\documentclass[manuscript]{acmart}

\renewcommand\footnotetextcopyrightpermission[1]{} 
\AtBeginDocument{%
  \providecommand\BibTeX{{%
    \normalfont B\kern-0.5em{\scshape i\kern-0.25em b}\kern-0.8em\TeX}}}

\setcopyright{acmcopyright}
\copyrightyear{2022}
\acmYear{2022}
\acmDOI{}

\acmConference[]
\acmPrice{}
\acmISBN{}

\usepackage{booktabs}
\usepackage{multirow}

\begin{document}

\title[The Impact of Feature Quantity on Recommendation Algorithm Performance]{The Impact of Feature Quantity on Recommendation Algorithm Performance: A Movielens-100K Case Study}

\pagestyle{plain}

\author{Lukas Wegmeth}
\email{lukas.wegmeth@uni-siegen.de}
\affiliation{%
  \institution{University of Siegen}
   \country{Germany}
}


\begin{abstract}
    Recent model-based Recommender Systems (RecSys) algorithms emphasize on the use of features, also called side information, in their design similar to algorithms in Machine Learning (ML). In contrast, some of the most popular and traditional algorithms for RecSys solely focus on a given user-item-rating relation without including side information. An important category of these is matrix factorization-based algorithms, e.g., Singular Value Decomposition and Alternating Least Squares, which are known to have high performance on RecSys data sets. The goal of this case study is to provide a performance comparison and assessment of RecSys and ML algorithms when side information is included. We chose the Movielens-100K data set since it is a standard for comparing RecSys algorithms. We compared six different feature sets with varying quantities of features which were generated from the baseline data and evaluated on a total of 19 RecSys algorithms, baseline ML algorithms, Automated Machine Learning (AutoML) pipelines, and state-of-the-art RecSys algorithms that incorporate side information. The results show that additional features benefit all algorithms we evaluated. However, the correlation between feature quantity and performance is not monotonous for AutoML and RecSys. In these categories, an analysis of feature importance revealed that the quality of features matters more than quantity. Throughout our experiments, the average performance on the feature set with the lowest number of features is $\sim$6\% worse compared to that with the highest in terms of the Root Mean Squared Error. An interesting observation is that AutoML outperforms matrix factorization-based RecSys algorithms when additional features are used. Almost all algorithms that can include side information have higher performance when using the highest quantity of features. In the other cases, the performance difference is negligible (<1\%). The results show a clear positive trend for the effect of feature quantity as well as the important effects of feature quality on the evaluated algorithms.
\end{abstract}

\ccsdesc[500]{Information systems~Recommender systems}

\keywords{feature engineering, recommender systems, automated machine learning, case study}

\maketitle

\section{Introduction}
Matrix factorization-based Recommender System (RecSys) algorithms are specialized for predicting missing entries, e.g., ratings, in sparsely filled user-item matrices. A handful of often-used benchmark data sets exist \cite{4688070, 10.1145/963770.963772}, which represent such a RecSys task. Interestingly enough, some of these data sets include side information, also called features, that are not used by the aforementioned RecSys algorithms. Instead, the data is directly reduced to a sparse user-item matrix, ignoring any side information. Machine Learning (ML) algorithms, in contrast, are fairly broad in their applications and usually profit from the availability of additional, meaningful features \cite{SUGUMARAN20114088, BLUM1997245, khalid2014survey}. As an extension, Automated Machine Learning (AutoML) techniques further increase ML performance through automated algorithm selection and hyperparameter optimization. Furthermore, the same tasks that RecSys algorithms intend to solve, can generally also be solved by (Auto)ML algorithms. Recent advances in RecSys have led to more sophisticated model-based algorithms, with the likes of Factorization Machines \cite{rendle2010factorization} and especially the latest Deep Neural Networks (DNN), that can incorporate side information and are more similar to their ML relatives \cite{cheng2016wide, zhou2018deep, 10.1145/2988450.2988454, GRIDACH202023, 10.1145/2959100.2959190}. However, it is problematic that the performance gap between (Auto)ML and RecSys has not been explicitly researched before.

Feature engineering is a broad topic that is well documented and researched due to its positive influences on ML \cite{zheng2018feature, dong2018feature, khalid2014survey, chandrashekar2014survey, seger2018investigation}. It summarizes many techniques for feature processing that mainly intend to increase prediction performance. Today, such feature engineering techniques are standard for many ML pipelines. Among those techniques is the curation of features by selection and extraction. Additional real-world features, as well as features extracted from existing data, are often beneficial to a model's performance. In this context, comparing the impact of feature quantity on the performance of RecSys and ML algorithms provides a meaningful indicator for the effects of feature engineering. However, we were not able to find previous comparative studies on the effect of feature quantity on RecSys algorithms.

Due to the aforementioned positive effects of feature engineering techniques in ML algorithms, we hypothesized that the same effects can also be shown for RecSys algorithms. Additionally, since a performance comparison between (Auto)ML and RecSys in terms of feature quantity is absent in the literature, the following two research questions arise. In a RecSys problem setting, how does feature quantity impact the performance:
\begin{itemize}
\item\emph{of ML, AutoML, and RecSys algorithms in general?}
\item\emph{of RecSys algorithms compared to (Auto)ML algorithms?}
\end{itemize}
We explore these questions through a case study on the Movielens-100K \cite{10.1145/2827872} data set. The code that produced the results reported in this paper is publicly available on our GitHub repository \footnote{\url{https://github.com/lukas-wegmeth/ML100K-Feature-Expansion}}.
 
\section{Methodology}
We evaluated 19 algorithms from nine libraries (Table \ref{algorithm_overview}) on six feature sets generated from the Movielens-100K \cite{10.1145/2827872} data set (Table \ref{feature_set_table}). 

\begin{table}[ht]
\centering
\caption{An overview of the evaluated algorithms and their category. We evaluated the algorithms as implemented in their respective library. Importantly, it is shown whether the algorithm can incorporate side information and how their hyperparameters were tuned.}
\label{algorithm_overview}
\resizebox{0.8\textwidth}{!}{%
\begin{tabular}{@{}|l|l|l|c|c|@{}}
\toprule
Category & Library & Algorithm & \multicolumn{1}{l|}{Uses side information} & \multicolumn{1}{l|}{Hyperparameter tuning} \\ \midrule
\multirow{1}{*}{Baseline} & \multirow{5}{*}{Scikit-Learn \cite{scikit-learn}} & Constant Predictor & - & - \\ \cmidrule(l){1-1} \cmidrule(l){3-5} 
\multirow{5}{*}{Machine Learning} &  & Linear Regressor & X & - \\ \cmidrule(l){3-5} 
 &  & K Nearest Neighbors & X & SMAC3 \cite{lindauer2021smac3} (200 runs) \\ \cmidrule(l){3-5} 
 &  & Random Forest Regressor & X & Random search: 20 iterations per fold \\ \cmidrule(l){3-5} 
 &  & Histogram Gradient Boosting Regressor & X & Random search: 20 iterations per fold \\ \cmidrule(l){2-5} 
 & XGBoost \cite{Chen:2016:XST:2939672.2939785} & Extreme Gradient Boosting Regressor & X & - \\ \midrule
\multirow{3}{*}{AutoML} & Auto-Sklearn \cite{feurer-neurips15a} & Best algorithm varies by fold & X & One hour search per fold \\ \cmidrule(l){2-5} 
 & H2O AutoML \cite{H2OAutoML20} & Best algorithm varies by fold & X & One hour search per fold \\ \cmidrule(l){2-5} 
 & FLAML \cite{DBLP:journals/corr/abs-1911-04706} & Best algorithm varies by fold & X & One hour search per fold \\ \midrule
\multirow{7}{*}{RecSys Matrix Factorization} & \multirow{3}{*}{Surprise \cite{Hug2020}} & SVD & - & - \\ \cmidrule(l){3-5} 
 &  & SVDpp & - & - \\ \cmidrule(l){3-5} 
 &  & KNNBaseline & - & - \\ \cmidrule(l){2-5} 
 & \multirow{3}{*}{Lenskit \cite{10.1145/3340531.3412778}} & User-User kNN collaborative filtering & - & - \\ \cmidrule(l){3-5} 
 &  & Item-Item kNN collaborative filtering & - & - \\ \cmidrule(l){3-5} 
 &  & Biased Alternating Least Squares & - & - \\ \cmidrule(l){2-5} 
 & \multirow{3}{*}{LibRecommender \cite{librec}} & SVDpp & - & Manually \\ \cmidrule(l){1-1} \cmidrule(l){3-5}
\multirow{3}{*}{RecSys Models}  &  & Wide \& Deep & X & SMAC3 \cite{lindauer2021smac3} (300 runs) \\ \cmidrule(l){3-5} 
 &  & Deep Interest Network & X & SMAC3 \cite{lindauer2021smac3} (100 runs) \\ \cmidrule(l){2-5} 
 & MyFM \cite{myfm} & Bayesian Factorization Machine & X & - \\ \bottomrule
\end{tabular}%
}
\end{table}

Movielens-100K \cite{10.1145/2827872} is one of the regularly used explicit feedback algorithm performance evaluation data sets in the RecSys community \citep{forouzandeh2021presentation, kuzelewska2014clustering, yang2016re, aljunid2020efficient}. The full data set consists of a table with user IDs and item IDs and their observed ratings where each rating has a timestamp. Additionally, each user ID and item ID contains a set of features specific to them. The user features are their age, gender, occupation, and (North American) ZIP code. The item features are their movie genre, title, release date, and IMDb URL. We solve the prediction of ratings as a regression task and measure and compare the performance of a given algorithm through the Root Mean Squared Error (RMSE).

To analyze the impact of the number of features on algorithms, we cut and/or enriched the features of the original data set and finally grouped them, resulting in six separate feature sets (Table \ref{feature_set_table}). The default feature set contains most of the basic features of the original data set. The idea here is to use as many of the original features as possible that require little to no further processing. For this reason, the item's title and IMDb URL as well as the user's ZIP codes were removed.

From the observed user-item-ratings relation many statistical features can be engineered. They can also be calculated in terms of the users and items separately. We calculated the following nine statistical features each in terms of the users and items: mean, median, mode, minimum, maximum, standard deviation, kurtosis, and skew. This provides a total of 18 additional features. Notably, these features were only calculated on the training set after splitting the data into a separate training set and test set to avoid leaking information about the training set into the test set. These engineered features should provide useful additional information to the algorithms at hand.

Finally, additional real-world data points can be added to the list of available features. For this, we chose the median and mean household income and population of a period as close as possible to the release date of the original data set. A set of these data points grouped by ZIP codes from the US from 2006 to 2010 \cite{income} is the earliest publicly available data that has a direct relation to the user features.

From various combinations of the aforementioned feature sets, we created a total of six combinations to perform the experiments on. An overview of the sets and their names that we refer to from here on is listed in Table \ref{feature_set_table}.

\begin{table}[ht]
\centering
\caption{An overview of the evaluated feature sets that we generated from the base Movielens-100K \cite{10.1145/2827872} data set either by cutting or enriching features. The columns denote the contained features in the named feature sets shown in the rows and also provide the total number of features.}
\label{feature_set_table}
\resizebox{0.8\textwidth}{!}{
    \begin{tabular}{@{}|c|c|c|c|c|c|@{}}
        \toprule
        \begin{tabular}[c]{@{}c@{}}
        features $\rightarrow$ \\ 
        set name $\downarrow$
        \end{tabular} &
        \multicolumn{1}{l|}{\begin{tabular}[c]{@{}l@{}}user ID\\ item ID\end{tabular}} &
        \multicolumn{1}{l|}{\begin{tabular}[c]{@{}l@{}}user stat. feat.\\ item stat. feat.\end{tabular}} &
        \multicolumn{1}{l|}{\begin{tabular}[c]{@{}l@{}}rating timestamp\\ user occupation\\ user age\\ user gender\\ movie genre\\ movie release date\end{tabular}} &
        \multicolumn{1}{l|}{\begin{tabular}[c]{@{}l@{}}user ZIP code\\ ZIP code income\\ ZIP code population\end{tabular}} &
        \multicolumn{1}{l|}{\begin{tabular}[c]{@{}l@{}}number of features \\categorical features count as one\end{tabular}} \\ \midrule
        stripped-no-stats            & X & - & - & - & 2\\ \midrule
        stripped-with-stats          & X & X & - & - & 20\\ \midrule
        basic-no-stats               & X & - & X & - & 8\\ \midrule
        basic-with-stats             & X & X & X & - & 26\\ \midrule
        feature-expansion-no-stats   & X & - & X & X & 11\\ \midrule
        feature-expansion-with-stats & X & X & X & X & 29\\ \bottomrule
    \end{tabular}}
\end{table}

We applied additional processing steps to some of the features to make them suitable for the evaluation, sometimes depending on the data set or automatically as an algorithm requires. Generally, we tried to stay as close to the original features as possible, making changes only where sensible or necessary. As a result, we did not treat the user ID and item ID as categorical features where applicable. The movie genre is provided as a categorical feature by default, which we did not change. However, the user's occupation is not provided as a categorical feature and we, therefore, transformed it into one. Movie release dates are provided in a date format and we converted them to a signed UNIX timestamp representation. The user's age and zip code are a special case. For the "basic" sets, we treated the age as an integer as it is provided in the original set. We removed the ZIP code since it contains some non-numerical entries and because we had no intention to filter ratings in these sets. For the "feature-expansion" sets, we divided the age by 18 to create five age categories and then treated the age as a categorical feature. We had to keep the ZIP code to add the mean and median household income and population features. Finally, we removed entries with ZIP codes that were not contained in the additional feature sets, which incurs a loss in observed ratings of 7.05\%. The remaining ZIP codes range from "00000" to "99999". To use them as a feature, we selected only the first digit of each ZIP code and then transformed it into a categorical feature. We chose to apply this processing step because the first digit in the ZIP code has a geographical meaning and therefore serves as an estimation for the residential area of each user.

To have an equal ground for algorithm comparisons, we applied some constraints. We performed all experiments on implementations in publicly available libraries to increase accessibility and reproducibility. The Movielens-100K \cite{10.1145/2827872} data set contains explicit ratings as integers ranging from one to five. Therefore, we only chose algorithms that can take explicit ratings as input and predict ratings in that same format. We evaluated the algorithms using five-fold cross-validation. Since one of the research questions is about a comparison of the performance of algorithms against each other, we performed hyperparameter tuning on algorithms that do not default to a tuned parameter setup for the Movielens-100K \cite{10.1145/2827872} data set. Depending on the algorithm, we tuned the hyperparameters either manually, with a random search, or by using SMAC3 \cite{lindauer2021smac3}, which is an all-purpose hyperparameter optimization tool. We set the time budget of AutoML tools to one hour for each fold. Table \ref{algorithm_overview} lists all libraries and algorithms and their categories, if they use side information or not, and how their hyperparameters were tuned.

\section{Results}

\begin{figure}[ht]
  \centering
   \begin{minipage}[t]{.48\textwidth}
   \caption{This plot shows the performance in terms of the RMSE of the algorithm categories as denoted in Table \ref{algorithm_overview} and plotted in Figure \ref{fig_benchmark} on the feature sets denoted in Table \ref{feature_set_table}. Each line plots the average RMSE of an algorithm category evaluated on each feature set represented by their number of features. The vertical lines and labels on the x-axis denote the collected data points. The first three data points do not include statistical features while the final three points are copies of the first three but include statistical features. The plot clearly shows a tendency for more features resulting in higher performance in general but also their effect on the categories themselves.}
    \label{fig_rmse_change}
    \includegraphics[width=\textwidth]{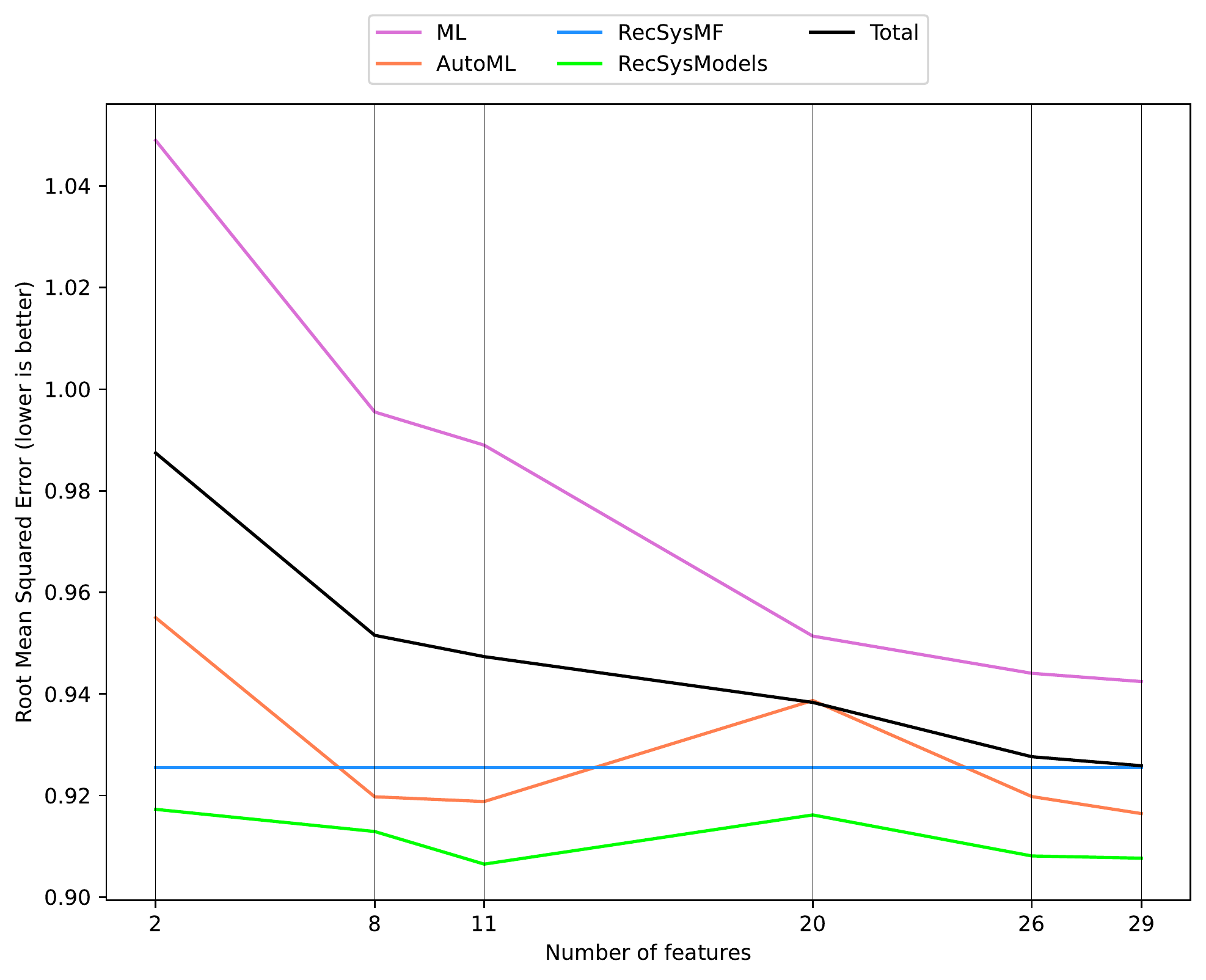}
  \end{minipage}
  \hfill
  \begin{minipage}[t]{.48\textwidth}
    \caption{The ordered Gini feature importance in percent. We evaluated these with a Random Forest Regressor that we fitted on the training data of the largest feature set. The chart shows the impact of each feature on the trained model. These important values are not necessarily true for the evaluated algorithms but provide a good estimation nonetheless. The feature names denote if they are an item or user feature. Statistical features are prefixed with 'i' and 'u' to denote this respectively.}
    \label{fig_feature_importance}
    \includegraphics[width=\textwidth]{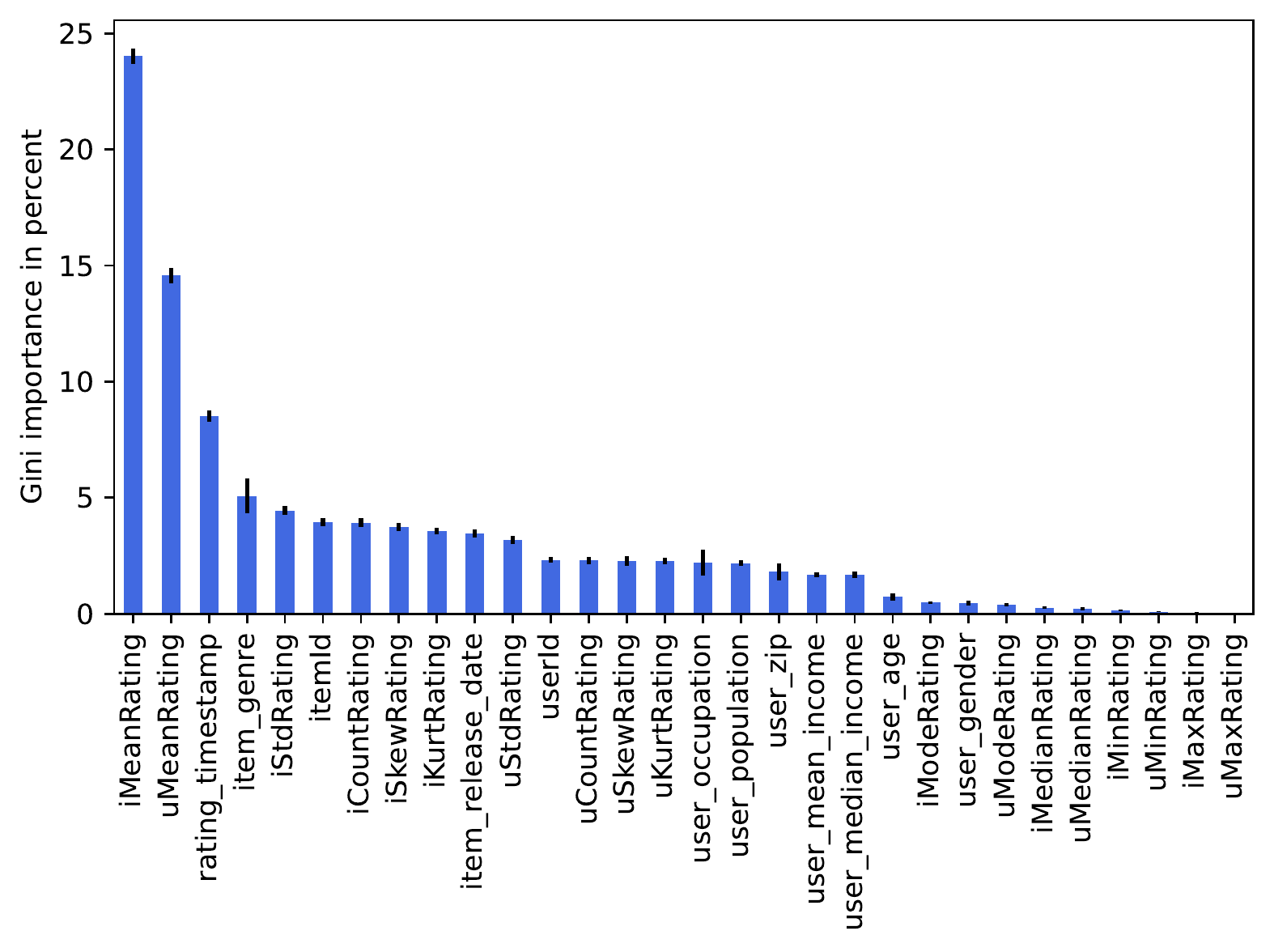}
  \end{minipage}
\end{figure}

Figure \ref{fig_rmse_change} aggregates the results gathered during the experiments and clearly shows that a higher feature quantity generally results in a lower RMSE. In particular, when comparing evaluations of the highest quantity of features with the lowest, the RMSE is \emph{~10\% lower for ML, ~4\% lower for AutoML, ~1\% lower for model-based RecSys, and ~6\% in total}.

\begin{figure}[ht]
    \centering
    \caption{This figure shows the detailed evaluation results that lead to the main conclusions presented in this paper. It shows the RMSE of the algorithms listed in Table \ref{algorithm_overview} on the feature sets listed in Table \ref{feature_set_table}. The bar chart is grouped by algorithms and the bars in each group are ordered by RMSE ascending. A lower RMSE is better. Algorithms that can not use additional features have the same performance on all feature sets. For these only, the "basic" feature set is plotted. These results are aggregated over algorithm categories in Figure \ref{fig_rmse_change}.}
    \label{fig_benchmark}
    \includegraphics[angle=90,width=\textwidth]{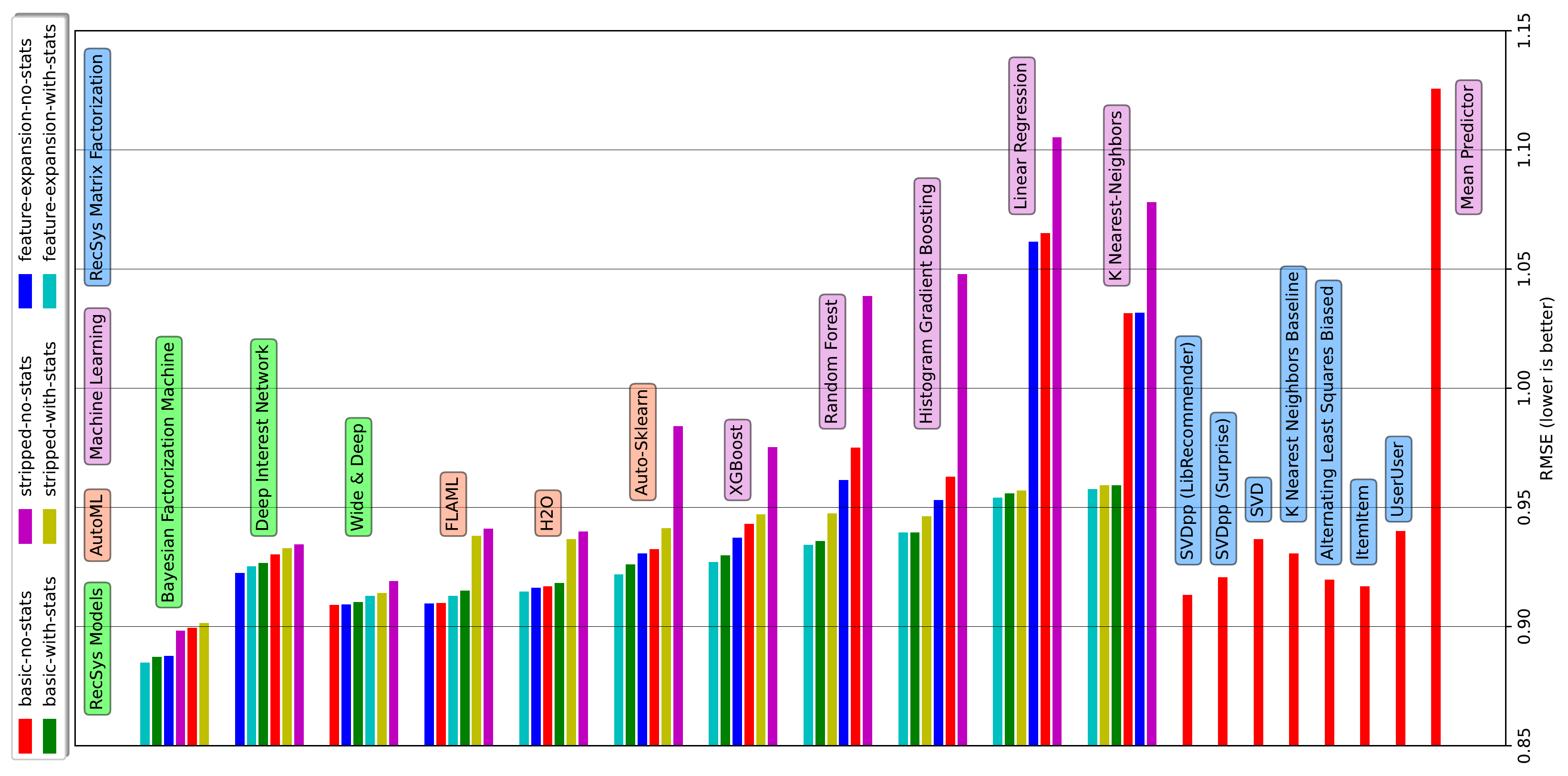}
\end{figure}

Importantly, the evaluation shows that AutoML outperforms the traditionally strong matrix factorization-based RecSys contenders in most of the evaluated feature sets. Even when only the "basic" feature sets are included, the RMSE is already ~1\% lower. Notably, however, there is an increase in RMSE between the feature sets with eleven features which is "feature-expansion-no-stats" and 20 features which is "stripped-with-stats". In these special cases, it appears that feature quantity alone can not significantly improve the algorithm's performance. One of the likely reasons is the feature importance seen in Figure \ref{fig_feature_importance}, which shows that the feature set with the higher feature quantity is missing some important features like the rating timestamp or item genre. 

Figure \ref{fig_benchmark} provides a detailed overview of all experiments. Interestingly, it shows that the performance difference and ranking in terms of the feature sets for RecSys and AutoML algorithms strongly varies by the algorithm which is in contrast to the ML algorithms where the ranking is mostly the same. Notably, the cross-validation procedure averages the results of multiple evaluations on an algorithm and the figures show these aggregations. However, the evaluations of an algorithm had only marginal performance differences (<2\%). Therefore, the reported averaged results shown here are relatively stable. Furthermore, we can observe that the performance of the different feature sets is divided into two groups for AutoML. One of the groups contains the "stripped" feature sets for which the search was not able to find a proper result even when provided with statistical features. In the other group, all of the other feature sets are tightly packed together, with a slight lead for the largest feature set, indicating the preference for a good combination and a higher quantity of real-world and statistical features.

Regarding the model-based RecSys algorithms, the DNN approaches that are Wide \& Deep and Deep Interest Network does not show a clear trend toward more features being favorable. One of the most interesting results, however, is the Bayesian Factorization Machine. Its performance increases clearly with feature quantity. Though a RecSys algorithm by nature, it could also be used to solve more general ML problems by design. It exhibits a mix of the behavior of ML and AutoML algorithms in its results which can be seen in Figure \ref{fig_benchmark}. The biggest notable difference is that its performance far exceeds that of any other compared algorithm on any feature set, making it a prime example of the capability of additional features for ML and RecSys alike.

ML algorithms strictly performed better with more features as expected. This evidences that the provided features have a good enough quality and distribution. In fact, the feature sets have an almost equal ranking order across all ML algorithms.

Overall, the results show a general favor towards a higher quantity of features in our study. The important caveat to this is the feature type and quality. As reported, due to feature importance, the relation between feature quantity and algorithm performance is not monotonous for AutoML and RecSys. The evaluated model-based RecSys algorithms performed best in this study, but our evaluation shows how even this gap can be closed through the introduction of additional features.

From the results, we conclude that the inclusion of statistical features is the most straightforward way to increase any algorithm's performance. Figure \ref{fig_feature_importance} clearly shows that the mean rating per user and item is especially effective. 

Only a relatively short time was invested to obtain and generate the additional features. However, these additional features had a positive impact on the performance of most of the tested algorithms. Given more care and time for feature engineering, the results may be significantly better. Next to the introduction of even more new features, there are many considerations for existing features and even for features that were introduced in this work. One of such is the fact that user IDs and item IDs, for example, are technically categorical features and should be treated as such, which was not always the case within the experiments due to constraints in the algorithms. There are also different ways to represent such categorical features which should be explored in this context. For example, we made the conscious decision to represent age as a categorical feature for one of the feature sets. These types of decisions have a potentially enormous impact on the performance of any algorithm and should be dealt with carefully.

\section{Future Work}

Our analysis of the feature importance as seen in Figure \ref{fig_feature_importance} clearly shows that some of the statistical features have especially high importance while other features have comparatively low importance. This is a good indication of the application of feature selection techniques which were out of the scope of this research.

The biggest challenge for directly profiting from the discoveries in this paper is likely to find suitable new features. Finding good data that supplies an existing data set is a hard task. However, it should be fairly easy to collect such feature data from the start when recording a new data set. Given our results, we encourage that future data set collection tasks include as many features as possible. Our work has shown that, for our scenario, if the goal is prediction performance, it is highly likely that a good selection of features in combination with a dedicated tuning will lead to better results. For future work on algorithm evaluations, we, therefore, propose that pipelines include feature engineering in terms of feature quantity because it may significantly improve results.

Finally, due to the rise of DNN in RecSys, the effect of additional features should also be studied extensively for this category of algorithms. The lack of available implementations and our self-imposed constraints within the research question led to only a small but nevertheless interesting peek at the effect of features on DNN RecSys algorithms. And given the observed results, there should be interest in investing more work in the future to understand the full effect of features on DNN in RecSys.

\bibliographystyle{ACM-Reference-Format}
\bibliography{bib}

\appendix

\end{document}